# EDGE CHANNEL DOMINATED MAGNETOTRANSPORT IN PbTe WIDE PARABOLIC QUANTUM WELLS

J. OSWALD, G.SPAN, A. HOMER, G. HEIGL, P. GANITZER,

Institute of Physics, University of Leoben, A-8700 Leoben, Austria

E-mail: oswald@unileoben.ac.at

D.K. MAUDE, J.C. PORTAL

High Magnetic Field Laboratory, CNRS, BP 166 Grenoble, France

In PbTe wide parabolic quantum wells (WPQW) a plateau-like structure is observed in the Hall resistance, which corresponds to the Shubnikov-de Haas oscillations in the same manner as known from the quantum Hall effect. At the same time a non-local signal is observed which corresponds to the structure in $R_{xx}$ and $R_{xy}$. We find a striking correspondence between a standard quantum Hall system and this quasi 3D WPQW system.

## 1 Introduction

A study of the intermediate regime between 2-dimensional (2D) and 3-dimensional (3D) electronic systems has attracted a lot of interest in the past [1]. However, the realization of such so called "wide parabolic quantum wells" (WPQW) with a flat potential in the electron channel requires a parabolic bare potential, which is difficult to obtain on the basis of GaAs/Al$_x$Ga$_{1-x}$As heterostructures by MBE growth. WPQWs can be made much more easily using the n-i-p-i concept with Lead Telluride (PbTe). PbTe WPQWs have been recently realized successfully and investigated experimentally and theoretically [2, 3]. The main facts are as follows: The quantum Hall effect is suppressed and replaced by conductance fluctuations (CF) of universal amplitude $e^2/h$ [4,5,6]. The magnetotransport is dominated by edge channel conduction with strong, permanent back scattering, which gives rise to an overall linear increase of $R_{xx}$ with magnetic field [7]:

$$R_{xx} = p \cdot L \cdot R_{xy} \tag{1}$$

p denotes a back scattering parameter per unit of length for the edge electrons and L is the length between the voltage probes of the Hallbar. The observed CF are explained to result from fluctuations of this back scattering process, which becomes most dominant in the narrow contact arms [5,6]. According to Eqn.1 the superimposed bulk-Shubnikov-de Haas (SdH) oscillations in $R_{xx}$ must result from oscillations of the back scattering parameter p. For an experimental verification we used a WPQW with a high Fermilevel in the electron channel in order to get SdH oscillations also in the high magnetic field regime.

## 2 Experimental results

The experiments are performed with a standard lock-in technique like described in [2,4,5,6]

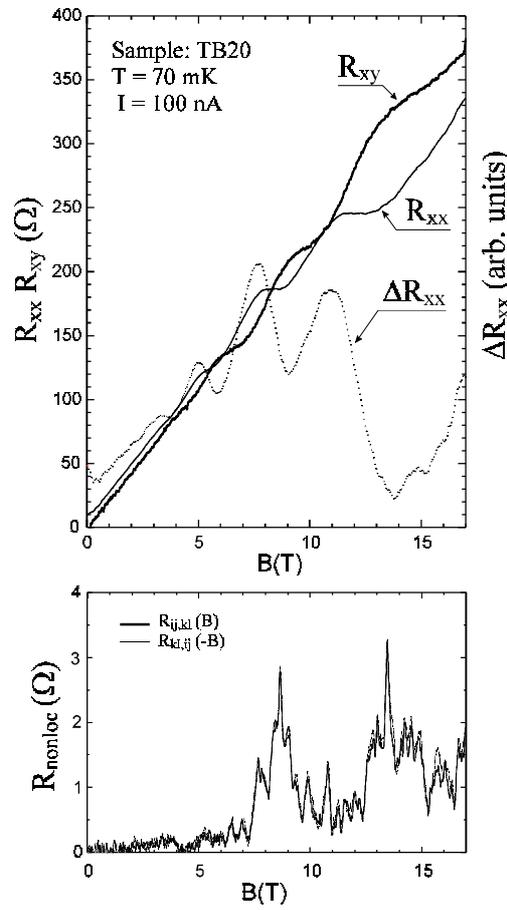

**Figure 1:** Magneto transport data of sample TB20. The upper diagram shows $R_{xx}$ and $R_{xy}$ data and the lower diagram shows the corresponding non-local data.

Figure 1 shows $R_{xx}$, $R_{xy}$ and non-local data. In addition, the $R_{xx}$ data are shown after subtracting the linear back ground ($\Delta R_{xx}$). SdH oscillations are observed up to B = 15 T and the resulting bulk density in the electron channel is $n_{3D} = 6 \times 10^{17} \text{cm}^{-3}$, which results in an effective electron channel width of about $d_n = 500 \text{nm}$.

### 3 Discussion

The basis for the discussion are the following experimental facts: (i)There are onsets of plateaus in $R_{xy}$ which correspond to minima in $\Delta R_{xx}$, (ii) the maxima in $\Delta R_{xx}$ appear in the regime of maximal slope in $R_{xy}$, (iii) the "plateaus" in $R_{xy}$ correspond to the bulk-SdH oscillations and therefore cannot be attributed to 2D-fillingfactors, (iv) the non-local signal including the superimposed fluctuation like structure fulfill exactly the Onsager-Casimir relation.

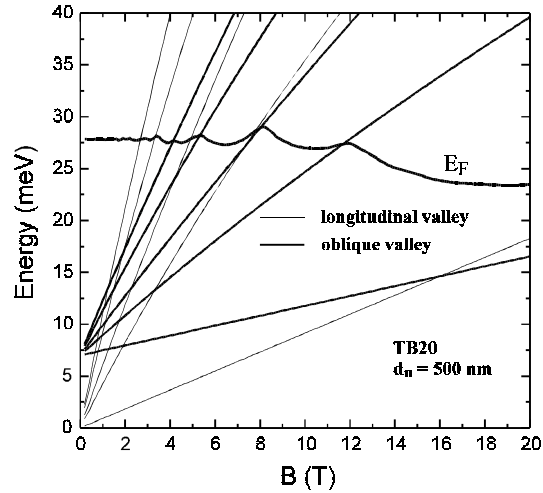

**Figure 2:** Bulk Landaulevels for the two different types of valleys and Fermienergy vs. magnetic field for B∥[111] and parameters according to sample TB20

Figure *2* shows a calculation of the LLs and the Fermienergy as a function of magnetic field for the parameters of the investigated sample. For the case of a confinement in [111] direction and a perpendicular magnetic field there are 2 sets of Landau levels and two sets of subbands with significantly different Landau and subband splitting. The small confining mass can lead already to EC-conduction while the second one leads still to a 3D-like behavior [7]. In Fig.2 there can be found 3 dominating intersections of the Fermienergy with bulk LLs, which correspond very well to maxima of $\Delta R_{xx}$ in the experimental data of Figure 1. It is easily seen, that at magnetic fields, where $E_F$ approaches a bulk LL, there is a maximum in $\Delta R_{xx}$ and a maximal slope in $R_{xy}$. If $E_F$ is between bulk LLs, a minimum in $\Delta R_{xx}$ and an onset of a plateau in $R_{xy}$ appears. If we look at the non-local signal, we realize, that it is at the maximum if $\Delta R_{xx}$ is at the minimum. Based on Eqn.1, the minima in $\Delta R_{xx}$ indicate regimes where back-scattering is minimal and thus EC-related non-local transport can be most efficient, exactly consistent with the experimental observation.

All together we get a remarkable correspondence between our quasi 3D system and a standard QH-system: (i) In standard QH-systems EC-backscattering occurs if a 2D-LL crosses $E_F$ while it is maximal in the WPQWs if $E_F$ crosses a bulk-LL. (ii) The energy range between bulk-LLs in the WPQW corresponds to the energy range between LLs in the QH-system: While in this regime EC-backscattering is completely suppressed in the standard QH-system, it appears to be just significantly reduced in the WPQW-system. (iii) While the Fermilevel in 2D systems moves in a so called DOS-gap, it is pinned by localized states, which allows a continuous movement of the Fermilevel. In our WPQWs the role of the localized states is taken over by the bulk-like system (longitudinal valley), which, however, does not create localized states. Therefore the zeros in $R_{xx}$ are missing and a linear slope is superimposed on $R_{xx}$ instead.

## Acknowledgments

Support by FWF Austria under project P10510-NAW and through the TMR Programme of the European Community under contract ERBFMGECT950077.